\documentclass[aps,pre,preprint]{revtex4-1}

\usepackage{graphicx}
\usepackage{amsmath,amssymb}
\usepackage{psfrag}
\usepackage{color}

\newcommand{\kB}{k_{\scriptscriptstyle \rm B}}

\begin{document}

\title{Casimir-like forces in cooperative exclusion processes}

\author{Mauro Sellitto} 

\affiliation{Dipartimento di Ingegneria, Universit\`a degli Studi della
  Campania ``Luigi Vanvitelli'', Via Roma 29, 81031 Aversa,
  Italy. \\ The Abdus Salam International Centre for Theoretical
  Physics, Strada Costiera~11, 34151 Trieste, Italy.  }

\begin{abstract}
I show that cooperative exclusion processes with selective kinetic
constraints exhibit fluctuation-induced forces that can be attractive
or repulsive, depending on the density of boundary reservoirs, when
their density-dependent diffusion coefficient exhibits a minimum.  A
mean-field analysis based on a nonlinear diffusion equation provides an
estimation of the magnitude and sign of such a tunable Casimir-like force
and suggests its occurrence in interacting particle systems with a
diffusivity anomaly.
\end{abstract}

\maketitle

\date{\today}

Casimir forces, experienced by objects immersed in a fluctuating
medium with long-range correlations, are virtually present in all
fields of physics, from exotic states of matter such as evaporating
black holes, to the more ordinary liquid state structure (where they
are embodied as van der Waals forces) and critical
phenomena~\cite{Lamoreaux,dispersion,KaGo}.  There has been a
substantial theoretical and experimental effort to precisely
characterize those physical situations where such forces are repulsive
rather than attractive (see, for
example,~\cite{Armand,Capasso,Dietrich,Bechinger,Burton,Wilczek,Zhao}),
as initially found by Casimir in the ideal case of two parallel and
perfectly conducting metal plates. This possibility, predicted by the
Lifshitz theory when the dielectric properties of the plates and the
intervening medium are properly chosen, offers the opportunity of
engineering Casimir forces for the practical design of micro- and
nano-structured electro-mechanical devices and
materials~\cite{French,Rudi}. Parallel to these developments,
Casimir-like forces induced by {\em non-equilibrium} fluctuations have
received a growing attention in the past few
years~\cite{Sasa,Rami,Cat,Wolf,Andrea,Ted}.  Interest in this
extension of the Casimir effect lies in the intrinsic long-range
nature of correlations in nonequilibrium steady
states~\cite{Zarate,Dorfman}, which offers the advantage of requiring
no fine-tuning of the external control parameters.  Compared to their
quantum or critical counterpart these situations are relatively less
understood as the dynamical contribution to the force is more
difficult to access.

The paradigmatic case of driven diffusive systems coupled to particle
reservoirs at different density, $\rho_+$ and $\rho_-$, has been
recently addressed in Ref.~\cite{Aminov}.  The authors found the
remarkable result that the fluctuation-induced force between two
parallel slabs at distance $d$ is given, to the leading quadratic
order in $\Delta \rho = \rho_+ - \rho_-$, by the formula:
\begin{eqnarray}
  F \simeq \frac{\kB T (\Delta \rho)^2 }{24 \, d} P'' \left[ \left(
    \frac{ \rho}{ P'} \right)'' + \left( \frac{ \rho}{ P'}
    \frac{D'}{D} \right)' \right],
  \label{eq.fif}
\end{eqnarray}
where $D(\rho)$ is a density-dependent diffusion coefficient, $P$ is
the pressure, and the prime denotes the derivative with respect to the
particle density $\rho$, evaluated at $\rho_-$.
Eq.~\eqref{eq.fif} is intriguing because it suggests that the
fluctuation-induced force can be either attractive (negative) or
repulsive (positive), depending of the relative sign and the magnitude
of the two terms in the square brackets.  In particular, the second
term on the right-hand-side shows an explicit dependence on the
dynamics through the appearance of the diffusion coefficient (and its
first and second derivatives).  Unfortunately, computing the bulk
diffusion is feasible only under a very special circumstance, known as
the ``gradient condition'', according to which the current through a
lattice bond can be written as the difference (i.e., the discrete
gradient) of some local function.  In this case the dynamical
contribution to the Green-Kubo formula (represented by the spatial sum
of the time integral of current-current correlation) vanishes and the
bulk diffusion coefficient can be computed as a static
average~\cite{Spohn_rev,Sasada}. For non-gradient systems, such as
those considered here, there are multi-site interactions that make
this task very challenging (see, however~\cite{Teomy,Arita} for some
progress in this direction), and therefore, identifying a class of
systems for which the sign of the force can be reversed remains a
widely open problem.
By using a mean-field approximation, we show in this paper that the
fluctuation-induced force, in exclusion processes with a minimum in
the diffusion coefficient $D(\rho)$, can be made attractive or
repulsive when the density of the boundary reservoirs is near the
minimum of $D(\rho)$.

We consider {\em cooperative} exclusion processes in which a particle
can move only if it has certain specific numbers of vacant neighbours,
before and after the move.  This notion, that generalises the idea of
kinetic constrains typically used for modelling the slow relaxation of
glassy systems~\cite{RiSo,KoAn}, can be equivalently reformulated by
saying that the move of a particle is {\em not} allowed if the number
of vacant neighbours, before or after the move (not counting the
departure and target site), belongs to a certain subset ${\mathcal S}$
of non-negative integer lower than $c$, the lattice coordination
number. Cooperative exclusion processes with such selective kinetic
constraints are unambiguously identified by ${\mathcal S}$ and $c$.
One important feature of their dynamics is the microscopic time
reversibility, i.e., the detailed balance condition (as the constraint
acts in exactly the same way on the departure and target sites). Since
the only static interaction is the hard-core particle exclusion, the
equilibrium measure is that of a non-interacting lattice-gas and,
therefore, the entropy density is $s(\rho) = - \kB \left[ \rho \ln
  \rho (1-\rho) + \ln (1-\rho)\right]$. This means that for all
members of this family the pressure at particle density $\rho$ is:
\begin{eqnarray}
  P(\rho) & = & - \kB T \log(1-\rho).
  \label{eq.P}
\end{eqnarray} 
The interesting consequence is that, after a straightforward
manipulation of the terms in the square brackets of
Eq.~\eqref{eq.fif}, the force $F$ turns out to be repulsive if:
\begin{eqnarray}
  -2 + (1-2 \rho) \frac{D'}{D} + \rho(1-\rho) \frac{D''}{D}
  -\rho(1-\rho) \left(\frac{D'}{D} \right)^2 & > & 0.
  \label{eq.sign}
\end{eqnarray} 
The analysis of the sign of the various terms suggests that this
inequality holds if there is range of density in which
$D''>0$. Therefore, driven diffusive systems with a a minimum in the
diffusion coefficient are the most suitable candidates in which the
sign of the fluctuation-induced force can be reversed.
For some examples of physical systems displaying such
  a type of diffusivity anomaly, where the relaxation dynamics becomes
  first slower and then faster upon increasing isothermal compression,
  see~\cite{Ludo,Massimo}.

To estimate the diffusion coefficient of cooperative exclusion
processes we use a mean-field argument which neglects particle
correlations, according to which the average of the hopping
probability, which is a certain function of multi-site occupancy
variables, is approximated as a function of the average of the single
site occupancy variable (i.e., the particle density $\rho$).  This
gives~\cite{Se_CEP}:
\begin{eqnarray}
  D^{\scriptscriptstyle \rm NC}_{\scriptscriptstyle {\mathcal S}}(\rho)
  & = & \left[ 1 - \sum_{i \, \in \, {\mathcal S} } {c-1 \choose i}
    \rho^{c-1-i} (1-\rho)^i \right]^2,
  \label{eq:D_NC}
\end{eqnarray}
where the binomial terms account for the multiplicity of possible
configurations of particles and vacancies, that prevent hopping (with
the power 2 coming from the detailed balance condition).  
%
This mean-field or no-correlation (NC) approximation provides a
general upper bound on the actual diffusion coefficient and, although
rather crude, it is surprinsingly good and versatile in predicting
unsual convexity-change density profiles in boundary-driven
nonequilibrium steady-states~\cite{Se_CEP}.
Moreover, for sufficiently large system size, transverse local density
fluctuations are uncorrelated~\cite{Se_CEP}:
\begin{eqnarray}
  \langle \rho(x)^2 \rangle - \langle \rho(x) \rangle^2 & = &
  L^{-1} \langle \rho(x) \rangle (1 - \langle \rho(x) \rangle ) ,
\end{eqnarray}
with $L$ being the distance between the particle reservoirs. This
confirms that in the steady state the local pressure at position $x$
can be computed by means of Eq.~\eqref{eq.P} with $\rho = \langle
\rho(x) \rangle$.
To substantiate in more detail the above analysis of we now analyze
three specific cooperative exclusion processes on a square lattice,
$c=4$.

\begin{figure}[htbp]
\centerline{\input{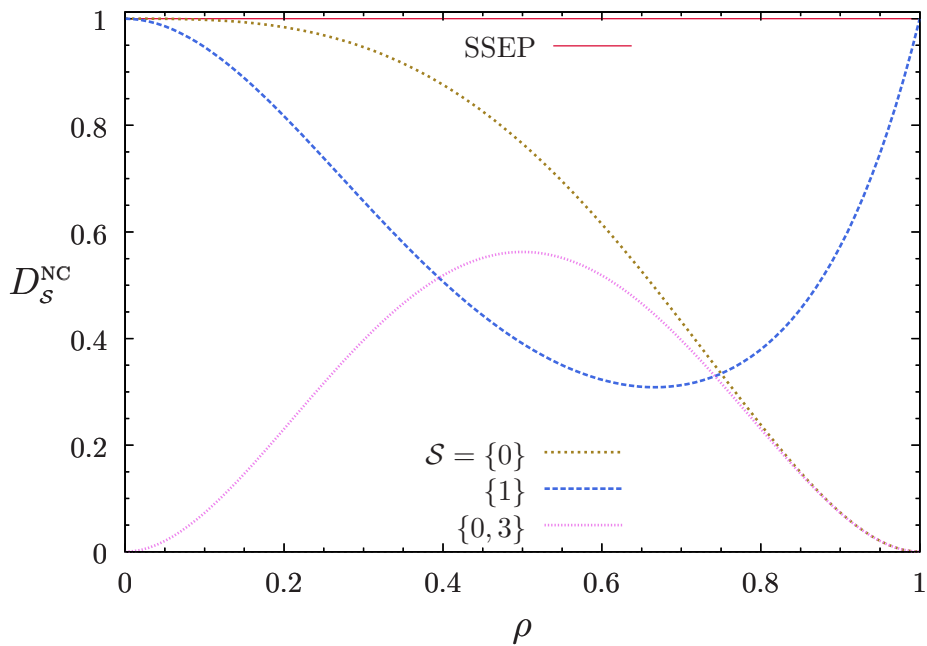}}
\caption{Bulk diffusion coefficient $D^{\scriptscriptstyle \rm
    NC}_{\mathcal S}$ in the no-correlation (NC) approximation for
  cooperative exclusion processes with selective constraint ${\mathcal
    S}$ on a square lattice.  The horizontal solid line corresponds to
  the exact value of the diffusion coefficient for the simple
  symmetric exclusion process (SSEP).}
\label{fig.D_NC}
\end{figure}

\begin{figure}[htbp]
\centerline{\input{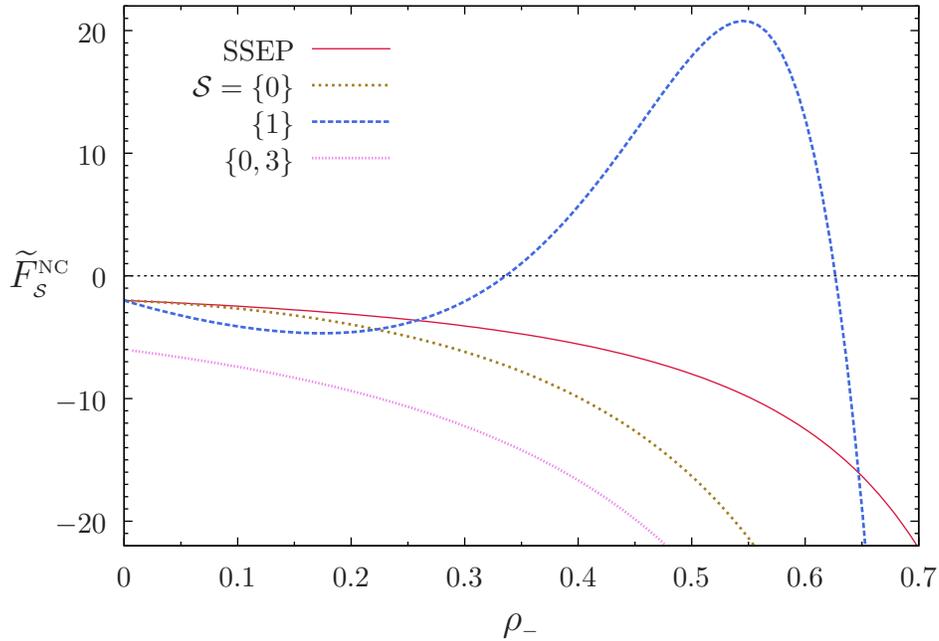}}
\caption{Fluctuation-induced force, $\widetilde{F}^{\scriptscriptstyle
    \rm NC}_{\mathcal S}$, in the no-correlation (NC) approximation vs
  particle reservoir density, $\rho_-$, for cooperative exclusion
  processes with selective constraint ${\mathcal S}$ on a square
  lattice. The exact value for the simple symmetric exclusion process
  (SSEP) is shown as a solid line for comparison.}
\label{fig.FIF}
\end{figure}

\begin{itemize}

\item ${\mathcal S} =\{0\}$. This is the Kob-Andersen model of glassy
  dynamics~\cite{KoAn} in which particle hopping to a vacant neighbour
  is not allowed if the particle has only one vacant neighbour, before
  or after the move. The dynamics is strongly cooperative and the
  diffusion coefficient steeply decreases to zero as $\rho \to 1$. It
  can be approximated as:
\begin{eqnarray}
  D^{\scriptscriptstyle \rm NC}_0(\rho) & = & \left( 1 - \rho^3 \right)^2 .
  \label{eq.D0}
\end{eqnarray}
When coupled to particle reservoirs at its ends this model generally
exhibits convex profiles~\cite{Teomy,Arita}.

\item ${\mathcal S} =\{0,3\}$. This is the particle-hole symmetric
  analog of the Kob-Andersen model.  Particle hopping to a vacant
  neighbour is forbidden if the particle has one or four vacant
  neighbours, before or after the move. The dynamics is strongly
  cooperative when $\rho$ tends to 0 or 1. The approximated diffusion
  coefficient is:
  \begin{eqnarray}
    D^{\scriptscriptstyle \rm NC}_{0,\,3}(\rho) & = & \left[ 1 -
      \rho^3 -(1-\rho)^3 \right]^2 ,
    \label{eq.D03}
  \end{eqnarray}
  an displays a maximum at $\rho^{\star}=1/2$.  When coupled to
  particle reservoirs at its ends the model shows convexity-change
  profiles depending on the reservoirs density~\cite{Se_CEP}.

\item ${\mathcal S}=\{1\}$. Particle hopping to a vacant neighbour is
  not allowed if the particle has two vacant neighbours, before or
  after the move.  This dynamics is moderately cooperative in an
  intermediate range of density as the kinetic constraint becomes
  ineffective at very low or high density.  The approximated diffusion
  coefficient is:
  \begin{eqnarray}
    D^{\scriptscriptstyle \rm NC}_1(\rho) & = & \left[ 1 - 3 \rho^2
      (1-\rho) \right]^2 ,
    \label{eq.D1}
  \end{eqnarray}
  and has a minimum at $\rho^{\star}=2/3$.  When coupled to particle
  reservoirs at its ends the model exhibits convexity-change profiles
  depending on the reservoirs density~\cite{Se_CEP}.
\end{itemize}
The approximated diffusion coefficient, $D^{\scriptscriptstyle \rm
  NC}_{\mathcal S}$, of the above exclusion processes is shown in
Fig.~\ref{fig.D_NC}.
With these ingredients one can now easily estimate the
fluctuation-induced force $F^{\scriptscriptstyle \rm NC}_{\mathcal
  S}$. In Fig.~\ref{fig.FIF} we show the results obtained for the
three cooperative exclusion processes
in terms of the rescaled force:
\begin{eqnarray}
  \widetilde{F}^{\scriptscriptstyle \rm NC}_{\mathcal S} =
  F^{\scriptscriptstyle \rm NC}_{\mathcal S} \frac{24 d}{ \kB T
    (\Delta \rho)^2}.
\end{eqnarray} 
We see that when the diffusion coefficient decreases monotonically,
${\mathcal S} =\{0\}$, the fluctuation-induced force is always
attractive and its strength becomes larger and larger as the reservoir
density $\rho_-$ increases.  This behaviour is qualitatively similar
to what has been observed in the simple exclusion
process~\cite{Aminov}, but it is enhanced by the larger dynamical
correlations needed for particle rearrangements in the presence of
kinetic constraints.  This interpretation is confirmed by the stronger
attraction observed for the case ${\mathcal S} =\{0,3\}$ which is
highly cooperative at both high and low density.
The most interesting and perhaps surprising case occurs when the
diffusion coefficient displays a minimum, ${\mathcal S} =\{1\}$. In
this case the fluctuation-induced force becomes repulsive in an
intermediate range of density around the mimimum of the diffusion
coefficient and attractive outside (so there are two crossover
densities at which the Casimir-like force vanishes).  A direct
inspection of Fig.~\ref{fig.FIF} shows that the maximum of the
repulsive force, which is attained at reservoir density $\rho_- \simeq
0.55$, is twice as strong as the attractive force in the simple
exclusion process (and equals in magnitude to the attractive force in
the Kob-Andersen model).  More importantly, the presence of a
repulsive force can be inferred from the convexity properties of
density profiles at different values of the density of boundary
reservoirs~\cite{Se_CEP}, as suggested by the relation
$x'(\rho)=D(\rho)$ (which follows from the expression of the
steady-state current $J=-D(\rho) \partial_x \rho$).  This macroscopic
static signature can be a useful diagnostic in the experimental or
numerical investigation of model systems when a detailed knowledge of
the bulk diffusion coefficient is lacking.
Since Eq.~\eqref{eq:D_NC} holds in the limit of vanishing dynamical
correlations, the results computed above should be considered as a
lower bound on the actual magnitude of the fluctuation-induced force
(therefore, one should observe forces of larger magnitude in
Montecarlo simulation of cooperative exclusion process). Although
improved estimations can be obtained by using more refined expression
of the diffusion coefficient with the systematic approaches developed
in~\cite{Teomy,Arita}, we expect that corrections are rather mild. In
particular, the prediction of the force sign and the location of the
crossover between the attractive and repulsive regime should be quite
accurate, as expected from numerical simulation of convexity-change
profiles in the nonequilibrium steady state~\cite{Se_CEP}.

It would be interesting to validate the above predictions by
Montecarlo simulations and to investigate how they are possibly
modified when small inclusions, rather than parallel slabs, are
considered.  The ability to control the strength and sign of these
forces should provide further opportunities for manipulating and
assembling particles in nonequilibrium condition on the soft matter
microscopic scales.  It is also tempting to speculate that they might
play a role in the quasi-isothermal cellular environment, where
cooperativity is the rule rather than an exception.






\begin{thebibliography}{50}

\bibitem{Lamoreaux} S.K. Lamoreaux, Phys. Today {\bf 60}, 40 (2007).

\bibitem{dispersion} S.Y. Buhmann, {\em Dispersion Forces I}, (Berlin:
  Springer 2012).

\bibitem{KaGo} M. Kardar and R. Golestanian, Rev. Mod. Phys. {\bf 71},
  1233 (1999).

\bibitem{Armand} A. Ajdari, L. Peliti, and J. Prost,
Phys. Rev. Lett. {\bf 66},1481 (1991).

\bibitem{Capasso}
J.N. Munday, F. Capasso, and V.A. Parsegian,
Nature {\bf 457}, 170 (2009)

\bibitem{Dietrich} C. Hertlein, L. Helden, A. Gambassi, S. Dietrich
  and C. Bechinger
Nature {\bf 451}, 172 (2008).

\bibitem{Bechinger} U. Nellen, L. Helden, and C. Bechinger,
Europhys. Lett. {\bf 88}, 26001 (2009).

\bibitem{Burton} J.-J. Li\'etor-Santos and J.C. Burton,
Soft Matter {\bf 13}, 1142 (2017).

\bibitem{Wilczek} Q.-D. Jiang and F. Wilczek,
Phys. Rev. B {\bf 99}, 125403 (2019).

\bibitem{Zhao} R. Zhao, L. Li, S. Yang, W. Bao, Y. Xia, P. Ashby,
  Y. Wang, and X. Zhang,
Science {\bf 364}, 984 (2019).

\bibitem{French} R. H. French et al., Rev. Mod. Phys. {\bf 82}, 1887
  (2010).

\bibitem{Rudi} L. M. Woods et al., Rev. Mod. Phys. {\bf 88}, 045003
  (2016).

\bibitem{Sasa} H. Wada and S. I. Sasa, Phys. Rev. E {\bf 67},
  065302(R) (2003).

\bibitem{Rami} A. Najafi and R. Golestanian, Europhys. Lett. {\bf 68},
  776 (2004).

\bibitem{Cat} C. Cattuto, R. Brito, U. Marini Bettolo Marconi,
  F. Nori, and R. Soto, Phys. Rev. Lett. {\bf 96}, 178001 (2006).

\bibitem{Wolf} M. R. Shaebani, J. Sarabadani, and D. E. Wolf,
  Phys. Rev.  Lett. {\bf 108}, 198001 (2012).

\bibitem{Andrea} A. Furukawa, A. Gambassi, S. Dietrich, and H. Tanaka,
  Phys. Rev. Lett. {\bf 111}, 055701 (2013).

\bibitem{Ted} T. R. Kirkpatrick, J. M. Ortiz de Z\'arate, and
  J. V. Sengers, Phys. Rev. Lett. {\bf 110}, 235902 (2013).

\bibitem{Zarate} J.M. Ortiz de Z\'arate and J.V. Sengers, {\em
  Hydrodynamic Fluctuations in Fluids and Fluid Mixtures} (Amsterdam:
  Elsevier, 2006).

\bibitem{Dorfman} J.R. Dorfman, T.R. Kirkpatrick and J.V. Sengers,
  Annu. Rev. Phys. Chem. {\bf 45}, 213 (1994).


\bibitem{Aminov} A. Aminov, Y. Kafri, and M. Kardar,
  Phys. Rev. Lett. {\bf 114}, 230602 (2015).

\bibitem{Spohn_rev} H. Spohn, {\em Large Scale Dynamics of Interacting
  Particles} (Heidelberg: Springer-Verlag, 1991).

\bibitem{Sasada} M. Sasada, Ann. Appl. Prob. {\bf 28}, 2727 (2018).

\bibitem{Teomy} E. Teomy and Y. Shokef, Phys. Rev. E {\bf 95}, 022124
  (2017).
\bibitem{Arita} C. Arita, P. L. Krapivsky, and K. Mallick, 
  J. Phys. A: Math. Theor. {\bf 51}, 125002 (2018).

\bibitem{RiSo} F. Ritort and P. Sollich, Adv. Phys. {\bf 52}, 219
  (2003).

\bibitem{KoAn} W. Kob and H.C. Andersen, Phys. Rev. E {\bf 48},
  4364 (1993).

\bibitem{Ludo} L. Berthier, A.J. Moreno, and G. Szamel, 
Phys. Rev. E {\bf 82}, 060501(R) (2010).

\bibitem{Massimo} M. Pica Ciamarra and P. Sollich, J. Phys.:
  Condens. Mat. {\bf 27}, 194128 (2015).

\bibitem{Se_CEP} M. Sellitto, Phys. Rev. E {\bf 100}, 040102(R)
  (2019).

\end{thebibliography}
\end{document}